\documentclass{pasj00}

\begin{document}
\SetRunningHead{H.Watanabe et al.}{Magnetic Structure of Umbral Dots}
\Received{200?/??/??}
\Accepted{200?/??/??}

\title{Magnetic Structure of Umbral Dots Observed with Hinode Solar Optical Telescope}

\author{
Hiroko \textsc{Watanabe}\altaffilmark{1},
Reizaburo \textsc{Kitai}\altaffilmark{1},
Kiyoshi \textsc{Ichimoto}\altaffilmark{1},
Yukio \textsc{Katsukawa}\altaffilmark{2}
 } 
 
\altaffiltext{1}{ Kwasan and Hida Observatory, Kyoto University, Kamitakara, Gifu 506-1314, Japan }
\altaffiltext{2}{ National Astronomical Observatory of Japan, Mitaka, Japan }

\email{watanabe@kwasan.kyoto-u.ac.jp}

\KeyWords{Sun: sunspots --- Sun: magnetic fields} 

\maketitle

%%%%%%%%%%%%%%%%%%%%%%%  Abstract %%%%%%%%%%%%%%%%%%%%%%%%%%%%%%
\begin{abstract}
High resolution and seeing-free spectroscopic observation of a decaying sunspot was done with the Solar Optical Telescope aboard {\it Hinode} satellite.  The target was NOAA 10944 located in the west side of the solar surface from March 2 to March 4, 2007.  The umbra included many umbral dots (UDs) with size of $\sim$300 km in continuum light.  
We report the magnetic structures and Doppler velocity fields around UDs, based on the Milne-Eddington inversion of the two iron absorption lines at 6302{\AA}. 

The histograms of magnetic field strength({\it B}), inclination angle({\it i}), and Doppler velocity({\it v}) of UDs showed a center-to-limb variation.  Observed at disk center, UDs had (i)slightly smaller field strength ($\Delta B=-$17 Gauss) and (ii)relative blue shifts ($\Delta v=$28 m s$^{-1}$) compared to their surroundings.  When the sunspot got close to the limb, UDs and their surroundings showed almost no difference in the magnetic and Doppler values.  This center-to-limb variation can be understood by the formation height difference in a cusp-shaped magnetized atmosphere around UDs, due to the weakly magnetized hot gas intrusion.  In addition, some UDs showed oscillatory light curves with multiple peaks around 10 min, which may indicate the presence of the oscillatory convection.  We discuss our results in the frameworks of two theoretical models, the monolithic model (Sch$\ddot{\textrm{u}}$ssler \& V$\ddot{\textrm{o}}$gler 2006) and the field-free intrusion model (Spruit \& Scharmer 2006).
\end{abstract}

%%%%%%%%%%%%%%%%%%%%%%%%  Introduction  %%%%%%%%%%%%%%%%%%%%%%%%%%

\section{Introduction}
The sunspot is one of the most prominent structures in the solar photosphere, although there are many related unsolved problems remaining even today.  One of them is the source of energy transport in sunspots.  It is known that the radiative energy alone is not sufficient for accounting for the observed brightness of sunspots, so another form of convective energy transport is necessary (Deinzer 1965).   The study of umbral dots (UDs), which are tiny bright points in the umbra, are essential for understanding the energy transport in sunspots, since UDs are considered to be a manifestation of the convection. 

Parker (1979) suggested in his "spaghetti" model, that UDs are the radiative signatures of the top parts of field-free convective plumes.  The field-free plumes intrude from below the visible surface into a gap between magnetic bundles in a cluster-type sunspot.  These plumes are accompanied by smaller magnetic field strength, substantial upflows within UDs, and a cusp-shaped magnetic structure (Spruit \& Scharmer 2006).  Another promising mechanism is the magneto-convection in a monolithic sunspot.  The monolithic model considers a sunspot as the aggregation of uniform vertically thin columns, and UDs as a natural result of the overstable oscillatory convection, which is a preferred mode just below the photosphere (Weiss et al. 2002; Sch$\ddot{\textrm{u}}$ssler \& V$\ddot{\textrm{o}}$gler 2006).  The monolithic model predicts smaller field strength, upflows in the center of UDs in addition to downflows at their boundaries, and a cusp-like structure.  
 
Few spectroscopic observations of UDs have been done so far, because of their tiny size (less than \timeform{0.5"}) and low brightness.  Only in recent days, some spectroscopic works have been published (Pahlke \& Wiehr 1990; Lites et al. 1991).  Wiehr \& Degenhardt (1993) observed UDs in the lines Fe 6843{\AA} and Ca 6103{\AA}, and found field strength reduction up to 20\% and flatter field inclination only in the lower layer.  Weaker field ($\sim$500 Gauss) with more horizontal orientation ($\sim$\timeform{10D}) in UDs was reported in Socas-Navarro et al. (2004).   

As for the Doppler velocity field, substantial upward velocity ($\sim$1 km s$^{-1}$) is observed in the lower photosphere.  However in the lines formed in the upper photosphere, no strong velocity field is related to individual UDs.  Rimmele (2004) found upflows in excess of 1 km s$^{-1}$ in C {\footnotesize I} 5380{\AA} (lower photosphere) line, while no strong upflow in Fe {\footnotesize I} 5576{\AA} (upper photosphere) line.  Socas-Navarro et al. (2004) also found upflows of $\sim$250 m s$^{-1}$ in Fe 6303.46{\AA} (lower photosphere) line, while no obvious upflow in Fe 6302.5{\AA} (upper photosphere) line.  Recently, Bharti et al. (2007a) found upward velocity on the order of 400 m s$^{-1}$, surrounded by narrow downflow regions with $\sim$300 m s$^{-1}$ in Fe {\footnotesize I} 5576{\AA} line.  

The spectro-polarimeter (SP) on board of the {\it Hinode} Solar Optical Telescope (SOT) (Tsuneta et al. 2008; Suematsu et al. 2008; Ichimoto et al. 2008; Shimizu et al. 2008) made it possible to observe diffuse UDs at the center of the umbra, with resolution limit of  \timeform{0.3"} in a highly stable condition.  We derived magnetic field strength, the orientation of the magnetic field, filling factor, and Doppler velocity using the Milne-Eddington inversion code.  In the following sections, we describe the details of the observation in {\S}2, analyze Stokes V area asymmetry in {\S}3, show the statistical results of magnetic and Doppler fields of UDs in {\S}4, and finally discuss and summarize our findings in {\S}5. 

%%%%%%%%%%%%%%%%%%%%%%  Observation %%%%%%%%%%%%%%%%%%%%%%%%%%%
\section{Observation}
The SP observation was performed from March 2 through March 4, 2007, in parallel with the acquisition of the filtergram data analyzed in Kitai et al. (2007).  The target was NOAA 10944 with $\alpha$-type sunspot in its decaying phase.  The sunspot invoked no flaring nor surging activity, and almost disintegrated on March 5.  The region was located in the west side of the solar surface.  The heliocentric coordinate of NOAA 10944 was (S\timeform{6D}, W\timeform{17D}) on March 2, (S\timeform{5D}, W\timeform{30D}) on March 3, and (S\timeform{6D}, W\timeform{43D}) on March 4. For more information, please refer to Kitai et al. (2007).  

With the SP, a Normal Map mode was carried out from 00:10 UT to 00:50 UT on the three consecutive days.  The Normal Map mode scans an area with an integration time of 4.8 s per slit position.  The observation covers the field-of-view (FOV) of \timeform{80"}$\times$\timeform{80"} with a polarimetric accuracy of 0.1\%.  The spatial pixel size was \timeform{0.159"} in slit direction and \timeform{0.147"} in step direction.  The spectral FOV covers two absorption lines of Fe {\footnotesize I} 6301.5{\AA} ($g_{eff}=$1.66) and Fe {\footnotesize I} 6302.5{\AA} ($g=$2.5).  The SSW routine {\it sp\_prep.pro} was applied for the purpose of dark subtraction and flat fielding.  

%%%%%%%%%%%%%%%%%%%%%%  Stokes V asymmetry  %%%%%%%%%%%%%%%%%%%%%%%%

\section{Stokes {\it V} area asymmetry}
\begin{figure}
  \begin{center}
    \FigureFile(80mm,65mm){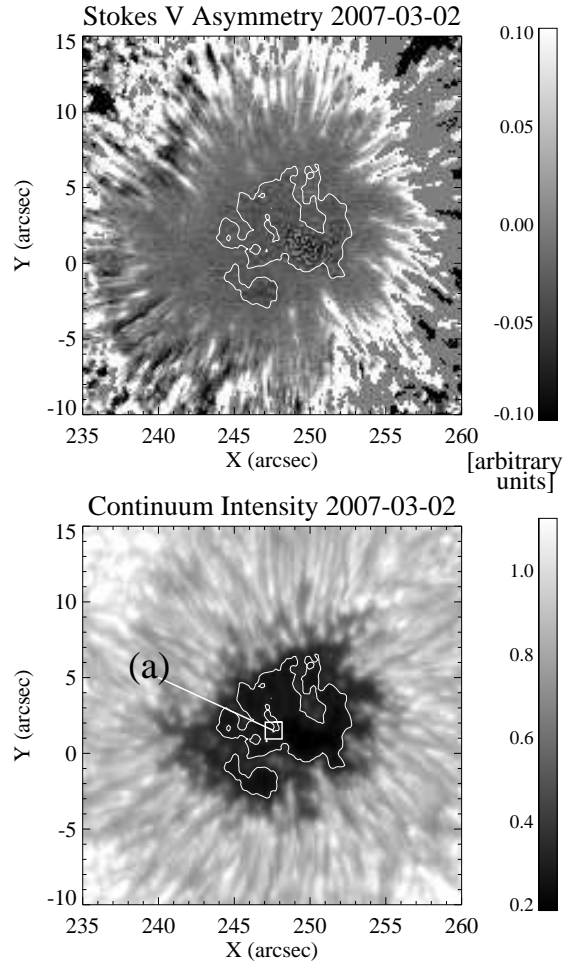}
  \end{center}
  \caption{Stokes {\it V} area asymmetry map ({\it top}) and continuum map at 6303{\AA} ({\it bottom}) on March 2.  The white curves show the smoothed contours of the continuum intensity at $I_{quiet} \times 0.3$.  Here $I_{quiet}$ is the average intensity of the quiet region.  The brightness of the continuum map is normalized by $I_{quiet}$.  The Stokes profiles at the position ({\it a}) are shown in Fig.\ref{fig:fitting}.}\label{fig:asym0302}
\end{figure}

Figure \ref{fig:asym0302} ({\it top}) shows the map of Stokes {\it V} area asymmetry in Fe {\footnotesize I} 6302.5{\AA} line on March 2.  The area asymmetry of Stokes {\it V} profiles provides an indication of a large gradient of field strength or line-of-sight velocity (Solanki \& Stenflo 1984, Stenflo \& Harvey 1985; Grossmann-Doerth et al. 1988, 1989; S{\'a}nchez Almeida \& Lites 1992).  We took the definition of  the Stokes {\it V} area asymmetry ($\delta A$) as 

\begin{equation}
\delta A =\frac{\int_{blue} |V| d\lambda - \int_{red} |V| d\lambda}{\int_{blue} |V| d\lambda + \int_{red} |V| d\lambda} \\
\end{equation}

\noindent where each integration is performed over the area of the blue or red lobe of {\it V} profiles.  In the dark core region (x=\timeform{250"}, y=\timeform{0"}), $\delta A$ is noisy, probably due to the blending of molecular lines at its cool temperature.  Except for the dark core, the umbra has a negligible value of $\delta A$.  This is because of the suppression of convective motions in the presence of strong magnetic field, which produces a small line-of-sight velocity in the umbra.  Morinaga et al. (2007) reported a smaller asymmetry in the center of the pore than its surrounding, which is consistent with our result.

%%%%%%%%%%%%%%%%%%%%%%  Magnetic Structure  %%%%%%%%%%%%%%%%%%%%%%%%

\section{Magnetic Structure and Doppler Velocity Distribution around UDs}

\subsection{Inversion}

\begin{figure}
  \begin{center}
    \FigureFile(80mm,65mm){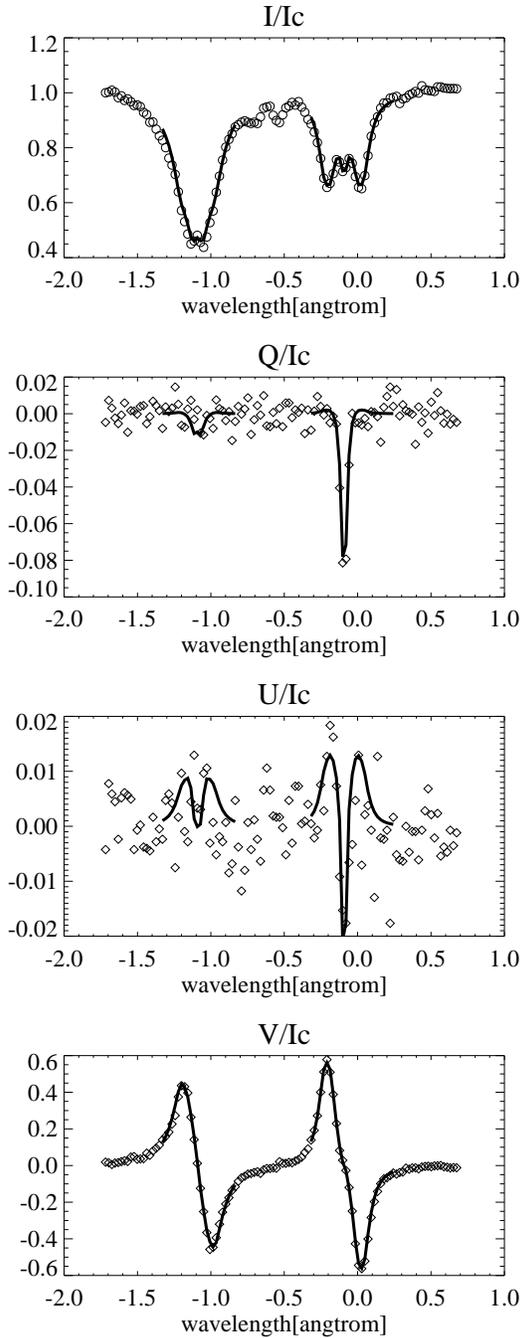}
  \end{center}
  \caption{Observed ({\it diamonds}) and best-fit Stokes {\it I, Q, U, V} profiles ({\it solid line}) by the inversion at the position ({\it a}) indicated in Fig.\ref{fig:asym0302}.  The profiles are normalized by the Stokes {\it I} intensity at continuum wavelength ({\it I}$_{c}$). }\label{fig:fitting}
\end{figure}

We applied a Milne-Eddington inversion code (Yokoyama et al. 2008, in preparation) to the Stokes spectra in.  As the Stokes {\it V} profile in the umbra has a negligible area asymmetry (\S{3}), it is reasonable to assume a Milne-Eddington atmosphere.  The best-fit UD profiles at the position ({\it a}) indicated in figure \ref{fig:asym0302} ({\it bottom}) are shown in figure \ref{fig:fitting}.  Figure \ref{fig:fitting} shows that Milne-Eddington inversion produces excellent fits to the observations.  The inversion code can derive 10 free parameters: three components of the magnetic field (strength, inclination, and azimuth), line-of-sight velocity, two parameters describing the linear dependence of the source function on optical depth, line strength, Doppler width, damping parameter, macroturbulent velocity, stray-light fraction, and a shift of the stray-light profile.  The \timeform{180D} ambiguity of the azimuth angle is determined by comparing with the potential fields calculated from the line-of-sight component of the magnetic field.  The stray-light profile is the averaged Stokes {\it I} profile over the regions where the maximum polarization degree ($p=\textrm{sqrt}(Q^2+U^2+V^2)/I$) along the line profile is larger than 0.2\%.  The stray-light represents the effect of a degradation of the polarization signals due to telescope diffraction and insufficient angular resolution (Orozco Suarez et al. 2007).  The magnetic filling factor, which represents the fraction of magnetized atmosphere, is computed as 1$-$(stray-light fraction).  
%The appearance of the maps are rotated as if the sunspot was seen from the top by the coordinate conversion, and the gaps in the rotated maps were interpolated by the nearest neighbor algorithm (Cover \& Hart 1967).  In this procedure, magnetic field inclination and field azimuth are converted with a planar approximation, that is, rotating the full FOV with a constant angle.  
The 2D maps and magnetic field vectors are converted to the local coordinate referring to the solar surface with the assumption that the solar surface is flat in our field of view, and, in the following, maps are presented as seen from the top.  
Doppler velocity is subtracted from the averaged Doppler velocity value inside the umbra.  The observed Doppler velocity field includes 3 min umbral oscillation, 5 min p-mode oscillation, and other instrumental effects.  We did not filter out these effects, since our interest is the local variations around UDs, which is easily distinguishable from them (umbral oscillation, p-mode, etc.).

We estimated the random error levels of the derived physical quantities from the standard deviation of the original map subtracted by a boxcar smoothed (width is \timeform{0.3"}x\timeform{0.3"}) map inside the umbra.  The smoothed width (\timeform{0.3"}x\timeform{0.3"}) is chosen to be narrower than the typical UD size, to calculate the fluctuation level contribution from other sources.  As a result, 1$\sigma$ error levels of field strength, field inclination, Doppler velocity, and filling factor are 13 Gauss, \timeform{0.7D},  10 m s$^{-1}$, and 0.02, respectively.

%%%%%%%%%%%%%%%%%%%%%%% Identification of UDs %%%%%%%%%%%%%%%%%%%%%%%

\subsection{Identification of UDs}

\begin{figure}
  \begin{center}
    \FigureFile(70mm,200mm){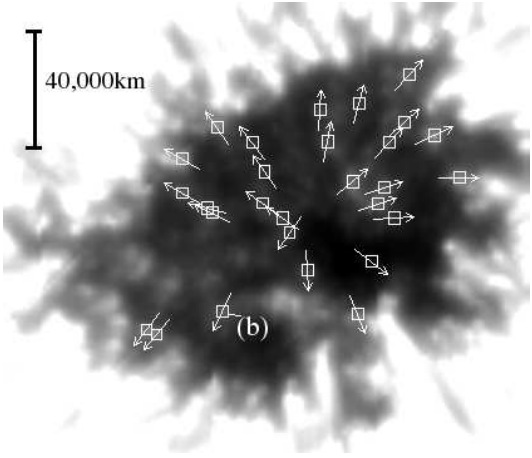}
  \end{center}
  \caption{Positions of 27 identified UDs on March 2.  The directions of the arrows mean the horizontal orientation of magnetic field at each position.  The plots at the position ({\it b}) are shown in Fig. 4.}\label{fig:identify}
\end{figure}

To identify UDs, we took the image segmentation method explained in Sobotka et al. (1997).  First, we made a boxcar smoothed (4x4 pixels) continuum map.  Then, the original continuum map was divided by the smoothed one, that is, {\it I}$_{c}$(original)/{\it I}$_{c}$(smoothed).  {\it I}$_{c}$ is the Stokes {\it I} intensity at continuum wavelength.  We set the empirical threshold value of 1.05 for identifying UDs, that is, an UD is more than 1.05 times brighter than its vicinity.  To avoid the influence of the statistical noise, UDs whose areas are less than 3 pixels are excluded from the analysis.  In this way, we identified 27 UDs on March 2 (Fig. \ref{fig:identify}), 35 on March 3, and 25 on March 4.  In total, 87 UDs are analyzed in this paper.

In Kitai et al. (2007), UDs are classified into 3 categories by their birth site, i.e., umbra, penumbra, and light-bridge origin.  The three categories of UDs show different characteristics with respect to their proper motion and temperature.   However, we don't take care of the subclasses in this paper, since we can not trace the birth site of each UD with our data sets.   

%%%%%%%%%%%%%%%%%%%%%%% Statistics over the three days %%%%%%%%%%%%%%%%%%%%%

\subsection{Results}  
\subsubsection{Statistics over the three days}  
\begin{figure}
  \begin{center}
    \FigureFile(70mm,20mm){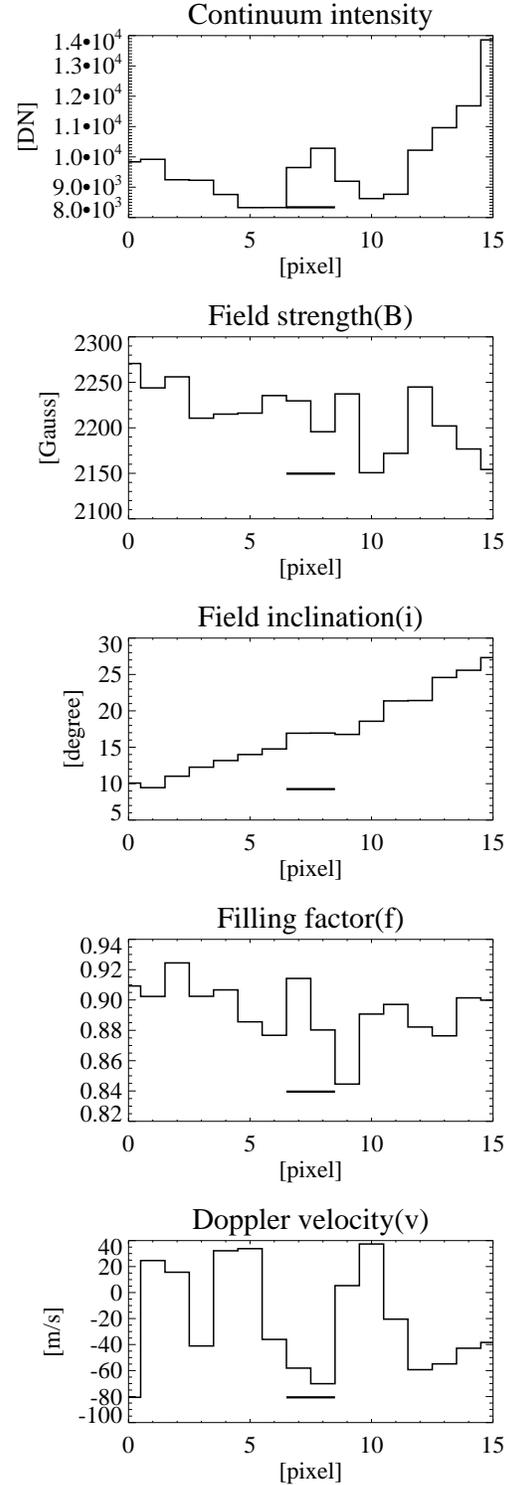}
  \end{center}
  \caption{Five plots around the UD indicated as ({\it b}) in figure \ref{fig:identify}.  From top to bottom, continuum intensity, field strength ({\it B}), field inclination ({\it i}), filling factor ({\it f}), and Doppler velocity ({\it v}) (positive means downflow).  A thick bar below each plots means the span of pixels which satisfy UD's conditions stated in \S4.2.  $x_{UD}=8, x_{BG}$=5 and 10.  $B_{UD}=2196$Gauss, $B_{BG}=2183$Gauss, $i_{UD}=$\timeform{17.0D}, $i_{BG}=$\timeform{16.3D}, $f_{UD}=0.880$, $f_{BG}=0.888$, $v_{UD}=-70.1$m s$^{-1}$, $v_{BG}=35.6$m s$^{-1}$.}\label{fig:plotex}
\end{figure}

We made spatial profiles of continuum intensity, field strength ({\it B}), field inclination ({\it i}), filling factor ({\it f}), and Doppler velocity ({\it v}) across each UD in the direction of the horizontal component of the magnetic field (shown in Fig. \ref{fig:identify} with arrows).  The plots for the UD ({\it b}) shown in figure \ref{fig:identify} are shown in figure \ref{fig:plotex}.  The horizontal axis covers the spatial length of  \timeform{2.2"} (15 pixels), which is long enough to cover the entire UD.  First, we decided the position of the UD ($x_{UD}$) and  its background ($x_{BG}$, 2 points at both side) by eye inspection of the local maximum and local minimums of the continuum intensity (Fig.\ref{fig:plotex} {\it top}).  Second, we calculated $\Delta F=F(x_{UD}) - F(x_{BG})$, where $F=B$, $i$, $f$, or $v$.  

\begin{figure}
  \begin{center}
    \FigureFile(80mm,20mm){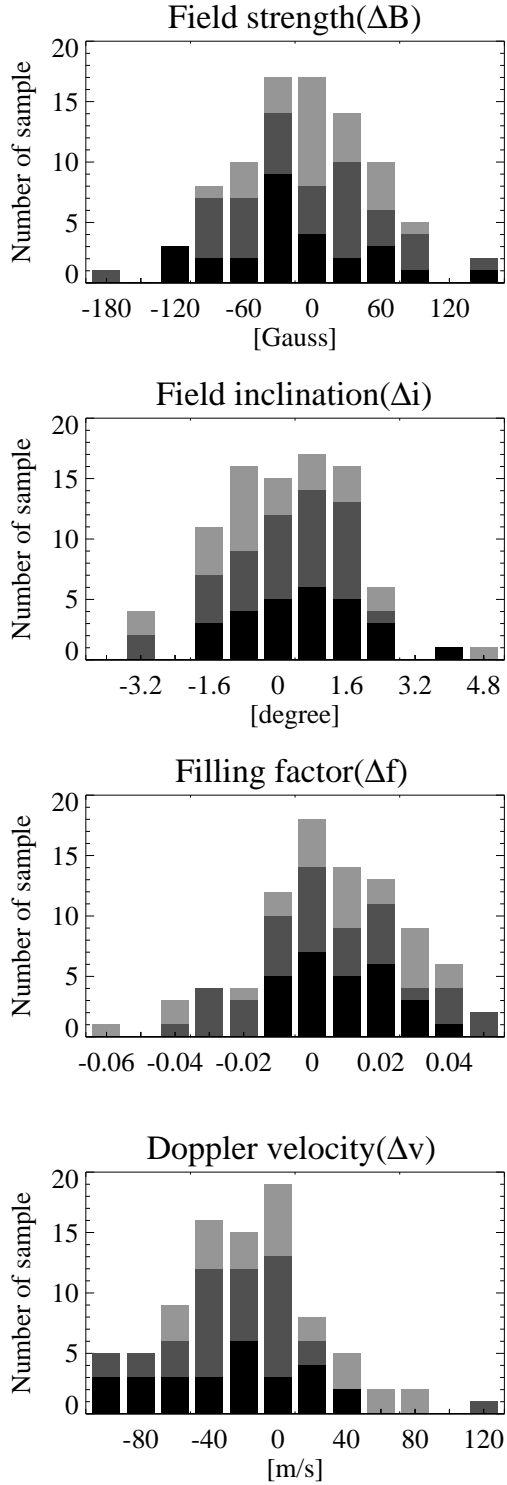}
  \end{center}
  \caption{Histograms of UD$-$BG differences.  The total sample number is 87.  From top, field strength ($\Delta B$), field inclination ($\Delta i$), filling factor ($\Delta f$), and Doppler velocity ($\Delta v$).  The black, gray, and light gray bars indicate the UDs on March 2, 3, and 4, respectively.}\label{fig:barplot}
\end{figure}

Figure \ref{fig:barplot} shows the histograms of the difference value ($\Delta F$) of four physical parameters.  
In the statistical average of all of the three day's data, UDs have relative blue shifts ($\Delta v_{average}=-$18 m s$^{-1}$).  As for the magnetic field, however, the statistical averages ($\Delta B_{average}=-$7 Gauss, $\Delta i_{average}=$\timeform{0.2D}, $\Delta f_{average}=$0.007) are smaller than the error levels, i.e., 13 Gauss, \timeform{0.7D}, and 0.02 respectively.  

\begin{figure}
  \begin{center}
    \FigureFile(75mm,65mm){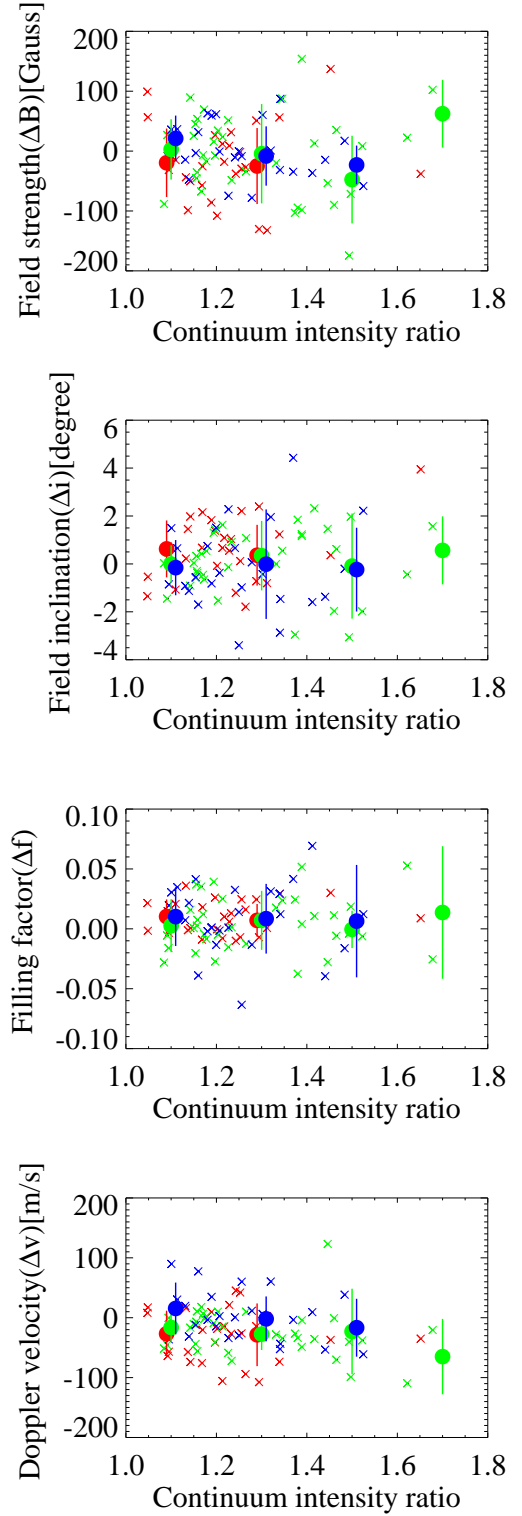}
  \end{center}
  \caption{Scatter plots of field strength ($\Delta B$), field inclination ($\Delta i$), filling factor ($\Delta f$), and Doppler velocity ($\Delta v$) against continuum intensity ratio UD/BG.  X signs are indicators of UDs on March 2 (red), March 3 (green), and March 4 (blue).  The circles show the average values in intensity ratio bin $\Delta$(UD/BG)=0.2 in each of the three observing days.  The colored solid lines show the standard deviation error bars. The negative value of $\Delta v$ means blue shift.}\label{fig:CPfill}
\end{figure}

In figure \ref{fig:CPfill} we show four scatter plots (from top, field strength $\Delta B$, field inclination $\Delta i$, filling factor $\Delta f$, and Doppler velocity $\Delta v$) against continuum intensity ratio UD/BG.  Larger continuum intensity ratio means a brighter UD.  Red (March 2), green (March 3), and blue (March 4) circles indicate the average values in intensity ratio bin $\Delta$(UD/BG)=0.2 with error bars showing the standard deviation.  There seems to be no correlation between continuum intensity ratio and $\Delta i$, and $\Delta f$ (the second and the third panels of Fig. \ref{fig:CPfill}).  On the other hand, $\Delta B$ and $\Delta v$ have weak correlations with continuum intensity ratio (the top and the bottom panels of Fig. \ref{fig:CPfill}).   That is, brighter UDs have weaker magnetic field, and larger blue shifts.\\

\subsubsection{Center-to-limb variation}

\begin{table}
  \caption{Center-to-limb variation of the averaged difference value UD$-$BG}\label{tbl:daily}
  \begin{center}
    \begin{tabular}{llll}
       & March 2 & March 3 & March 4 \\
       &S\timeform{6D},W\timeform{17D}&S\timeform{5D},W\timeform{30D}&S\timeform{6D},W\timeform{43D}\\\hline
      \hline
      Field strength & $-$17 & $-$6  & 1 \\
       (Gauss) $\Delta B$&   &   &  \\ \hline
      Field inclination & 0.6 & 0.1 & $-$0.1 \\
     (degree) $\Delta i$& &  &  \\ \hline
      Filling factor& 0.009 & 0.004 & 0.009 \\ 
       $\Delta f$&  &  &  \\ \hline
      Doppler velocity& $-$28 & $-$24 & 3 \\ 
      (m s$^{-1}$) $\Delta v$&  &  &  \\ \hline
    \end{tabular}
  \end{center}
\end{table}

As stated in \S4.3.1, UD's magnetic field, compared to their surroundings, does not show distinct variations in the statistical average of all of the three days data.  However, we found an interesting property in the daily statistics.  The difference values $\Delta F$ show center-to-limb variation, as listed in table 1.  $|\Delta B|$, $|\Delta i|$, and $|\Delta v|$ get smaller values as days go on, apart from $\Delta f$.  The sunspot was the closest to disk center on March 2.  Observed near disk center, UDs show smaller field strength, larger field inclination (more horizontal), and relative blue shifts.  On March 4, the sunspot was distant from the disk center.  In this case, UDs and their BG show almost no difference in magnetic and Doppler field.  As for the Doppler velocity, the projection effect may partly contribute to this result, since we detected the line-of-sight components.

\section{Discussion and Summary}

Our analysis revealed the magnetic structure and the velocity field distribution around UDs with seeing-free, high-sensitive {\it Hinode} SP observation.  The main results are as follows:
\begin{enumerate}
 \item The Stokes V profiles of UDs are virtually symmetric. 
 \item In the statistical average of all of the three days data, UDs do not produce distinct variations from the surroundings in their magnetic conditions, while Doppler velocity shows effective blue shifts ($-$18 m s$^{-1}$).
 \item The filling factor shows no difference on UDs.
 \item There are weak positive correlations between bright UDs and weaker field inclination, and between bright UDs and relative blue shifts.
 \item The local differences of field strength, field inclination, and Doppler velocity on UDs show center-to-limb variation.
\end{enumerate}

\subsection{Fe {\footnotesize I} formation height}

\begin{figure}
  \begin{center}
    \FigureFile(80mm,65mm){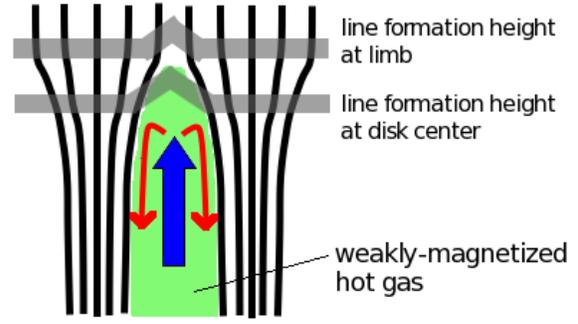}
  \end{center}
  \caption{Schematic figure of the cusp-shaped UD magnetic field lines and the Fe {\footnotesize I} line formation height.  The black solid lines are magnetic field lines.  The central green part corresponds to a weakly-magnetized hot gas, i.e., an UD.  Blue and red arrows indicate upflow and downflow, respectively.  The downflows, however, are an imaginary picture because we could not find them.  The two gray thick lines indicate the line formation heights at disk center (lower) and at limb (upper).  See text for further information.}\label{fig:UDmodel}
\end{figure}

According to Tritschler \& Schmidt (1997), the formation height of the Fe {\footnotesize I} 6302.5{\AA} in umbrae is 180 km higher than the height of continuum optical depth $\tau_{C}=1$ at 500nm for the line core, and the formation height above an UD is 130 km higher than $\tau_{C}=1$.  In addition, the continuum level of the UD is found to be shifted to $\sim$100 km higher layers (Degenhardt \& Lites 1993).  But these values are estimated at disk center, where we can look into the deepest layer.  When the target gets close to the limb, the formation height of spectral lines gets higher and higher.  This may explain the center-to-limb variation on UDs we've found in \S4.3.2.  The schematic view of an UD that accounts for the center-to-limb variation of magnetic and velocity fields is shown in figure \ref{fig:UDmodel}.  When we observe an UD at disk center, weakly-magnetized hot gas can reach the formation height of Fe {\footnotesize I}.  The observation reveals smaller field strength ($\Delta B=-17$Gauss), flatter field inclination ($\Delta i=$\timeform{0.6D}), and relative blue shifts ($\Delta v=-28$m s$^{-1}$), though field inclination difference is less than the error fluctuation level.  When we observe an UD far from disk center, the line formation height is higher than the UD occurring site.  The observation supports this interpretation because we could not find large differences between UDs and their BG on March 4.  Our results directly support the model that the UDs are formed at deep photosphere, and have a cusp-shaped magnetic field.  As for the downflows showed by red arrows in figure \ref{fig:UDmodel}, the discussion is made in \S5.3. 

There is a possibility that this center-to-limb variation is due to the evolutional phase difference of UDs in a decaying sunspot.  No one has examined the change of UD characteristics in a developing, mature, and decaying sunspot as far as we know.  Of course, the actual cause of this variation can be the mixture of the two, that is, the formation height difference and the evolutional phase difference.        

\subsection{filling factor}
As stated in \S4.1, the filling factor is computed as 1$-$(stray-light fraction).  In our inversion, the stray-light profile is the averaged Stokes {\it I} profile over the regions where the maximum polarization degree along the line profile is larger than 0.2\%.  According to the limb observation of SP performed on March 16, 2007, the fraction of the scattered light in the continuum wavelength was 2\% at a few arcsec away from the limb.  On the other hand, the average stray-light fraction inside the umbra was $\sim$10\%.  Thus the observed profile is considered to be composed of three components: unpolarized light due to telescope diffraction ($\sim$2\%), unpolarized light coming from the field-free atmosphere in the umbra, and polarized light coming from the magnetized gas.  The filling factor indicates the fraction of the polarized light. 

Assume that UDs are the penetration of field-free hot gas (Socas-Navarro et al. 2004; Spruit \& Scharmer 2006) and the penetration reaches the formation height of Fe {\footnotesize I} 6302{\AA}, the detection of small filling factor on UDs are predicted.  However, we could not find a decrease of the filling factor on UDs (table 1 and Fig. \ref{fig:CPfill}).  Higher resolution observation is strongly needed for further discussion.

\subsection{Comparison with sunspot models}

\subsubsection{The monolithic model}
A three dimensional simulation of UD phenomena was recently done by Sch$\ddot{\textrm{u}}$ssler \& V$\ddot{\textrm{o}}$gler (2006).  They explained UDs as a natural result of convection in a strong, initially monolithic magnetic field.  In their simulation, most of the UDs had an elongated form with a central or threefold dark lane which separates the UD into two or three parts.  At the end points of the dark lane there exist downflow patches.  

This picture is rather different from our observational impression.  We observed almost circular UDs and no dark lanes inside them with Fe {\footnotesize I} continuum map.  This is also the case with the blue/green continuum images obtained with {\it Hinode} SOT with spatial resolution \timeform{0.2"} (Kitai et al. 2007).  However in recent days there has ever some evidence of dark lanes with high resolution data, for example, Bharti et al. (2007b) with G-band filtergrams using {\it Hinode} SOT.  Thus far no one, including us, has succeeded in finding localized downflow patches at the end points of dark lanes (see Fig. 1 in Sch$\ddot{\textrm{u}}$ssler \& V$\ddot{\textrm{o}}$gler 2006).  There are two possibilities why we could not find localized downflow patches:  One is that, the downflow patches may be too small to be detected with {\it Hinode}'s resolution limit.  The other may be, because the continuum intensity (i.e., temperature) goes down when the gas flows downward, and the signal of the Stokes profile becomes too faint to be detected.  

\begin{figure}
  \begin{center}
    \FigureFile(80mm,65mm){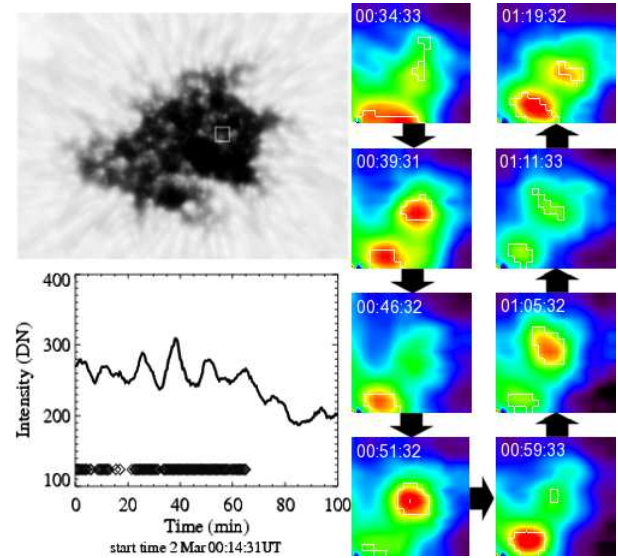}
  \end{center}
  \caption{Upper left: green continuum image on March 2, 2007.  The position of a periodic UD is marked with a white square.  Lower left: light curve of the center of the white square.  Below the light curve shown diamond signs which indicate more than 8\% brighter than the surroundings.  Right: Temporal change of zoom images of the white square region (pseudo-color display).  The time is written at upper left at each image in UT.}\label{fig:periplot}
\end{figure}

We found another important phenomenon of UDs.  It is already known empirically that some UDs occur and recur at the same location (Rimmele 1997).  With {\it Hinode}'s seeing-free condition, we got the light curve of successive UDs for about 3 hours using green continuum data.  Some UDs located in the center part of the umbra showed oscillatory light curves.  One example is shown in figure \ref{fig:periplot}.  The characteristic period of the oscillation is $\sim$10 minutes, and the successive emergence of UDs continues over $\sim$50 minutes.  This intensity oscillation may be supporting evidence for the monolithic model, because this model indicates that oscillatory convection is the preferred mode in the first few Mm depth below the umbral photosphere by linear stability analysis (Weiss et al. 1990).\\

\subsubsection{The field-free intrusion model}
Another plausible model of UDs is the field-free intrusion model.  This model was proposed by Parker (1979), in which he discussed that the magnetic field lines beneath the umbra are divided into many separate flux tube bundles, like "spaghetti".  The UDs would be a manifestation of field-free hot gas intrusion from below through the gaps of nonuniform magnetic field.  

A numerical study of the field-free intrusion model was done by Spruit \& Scharmer (2006) for penumbral grains and an observational study was done by Rimmele (2008).  They predict cusp-shaped magnetic field lines, smaller magnetic field, and upflow within the cusp.  These characteristics of UDs are almost the same as those of the monolithic model, apart from the localized downflow patches.  Field-free intrusion model also predicts downflow based on radiative cooling, but it's not concentrated on localized patches.  The weak correlation between dark UDs and downward motion, shown in figure \ref{fig:CPfill} may be evidence for downflow by radiative cooling.  As for the oscillatory light curve (Fig. \ref{fig:periplot}), however, the field-free intrusion model fails to predict such oscillation.  It would be very helpful if a future numerical simulation can make clear the possibility of brightness oscillation in the field-free hot gas surrounded by strong umbral magnetic field.

As was discussed in this paper, the oscillatory brightening of UDs seems to be a key phenomenon to reach a conclusion on the origin of UDs.  Detailed numerical studies and higher resolution observational, including the spectroscopic study of its temporal evolution on this phenomenon, are strongly needed in the near future.
\\

We are certain that we found many meaningful observational properties of magnetic field around UDs.  The reason why we succeed in deriving good correlations between UDs and the components of magnetic field, owes greatly to the really stable, sensitive, and high-resolutional observation performed by spectro-polarimetry on board of the {\it Hinode} Solar Optical Telescope.  Hinode is a Japanese mission developed and launched by ISAS/JAXA, with NAOJ as domestic partner and NASA and STFC (UK) as international partners. It is operated by these agencies in co-operation with ESA and NSC (Norway). 

%%%%%%%%%%%%%%%%%Acknowledgement%%%%%%%%%%%%%%%%%%%
\bigskip
The authors acknowledge T. Yokoyama, S. Morinaga, S. Morita, and all the staff and students of Kwasan and Hida Observatory whose comments were valuable for the improvement of the paper. The authors are supported by a grant-in-aid for the Global COE program "The Next Generation of Physics, Spun from Universality and Emergence" from the Ministry of Education, Culture, Sports, Science and Technology (MEXT) of Japan, and by the grant-in-aid for 'Creative Scientific Research The Basic Study of Space Weather Prediction' (17GS0208, PI: K. Shibata) from the Ministry of Education, Science, Sports, Technology, and Culture of Japan, and also partly supported by the grant-in-aid from the Japanese Ministry of Education, Culture, Sports, Science and Technology (No.19540474).

%%%%%%%%%%%%%%%%%%%%%%%%%%%%%%%%%%%%%%%

\end{document}